# HIGH-TEMPERATURE OXIDATION KINETICS OF ADDITIVELY MANUFACTURED NITIHF

**Hediyeh Dabbaghi[1], Mohammadreza Nematollahi, Keyvan Safaei Baghbaderani, Parisa Bayatimalayeri, Mohammad Elahinia**
Dynamic and smart systems lab
Toledo, OH

## ABSTRACT

*NiTi-based high-temperature shape memory alloys (HTSMAs) such as NiTiHf have been utilized in a broad range of applications due to their high strength and work output, as well as, their ability to increase the transformation temperatures (TTs). Recently, additive manufacturing techniques (AM) have been widely used to fabricate complex shape memory alloy components without any major modifications or tooling and has paved the way to tailor the manufacturing and fabrications of microstructure and critical properties of their final parts. NiTi alloys properties such as transformation temperatures can be significantly altered due to oxidation, which can occur during the manufacturing process or post-processing. In this work, the oxidation behavior of Ni-rich NiTi20Hf shape memory alloys, which was fabricated by the selective laser melting (SLM) method, is evaluated. Thermogravimetric analysis (TGA) is used to assess the kinetic behavior of the oxidation at different temperature ranges of 500, 700, and 900 °C for 20 hours in the air. After oxidation, to evaluate the microstructure and chemical composition X-ray diffraction (XRD), scanning electron microscopy (SEM), and energy-dispersive X-ray spectroscopy (EDS) was conducted. The isothermal oxidation kinetics of conventional NiTi20Hf alloys were studied, and the results were compared to AM samples. Results show a two-stage oxidation rate at which oxidation increased with the high rate at the initial stage. As the oxidation time increased, the oxidation rate gradually decreased. The oxidation behavior of NiTiHf alloys initially obeyed logarithmic rate law and then followed by parabolic rate law. SEM results showed the formation of a multi-layered oxide scale, including $TiO_2$, $NiTiO_3$, and Hf oxide.*

## 1. INTRODUCTION

Shape memory alloys (SMAs) such as binary NiTi have attracted much attention in various applications for their useful properties [1,2], such as superelasticity and shape memory effect. However, their low transformation temperatures (TTs) limit their applications in many industries that require high operating temperatures such as aerospace and automotive. In recent years, high-temperature shape memory alloys (HTSMAs) have been introduced by adding the third element to the binary NiTi alloys, which affects the TTs and other mechanical and thermomechanical properties of alloys. Among these elements such as Pd, Pt, Au, Zr, and Hf, NiTiHf is considered as one of the most effective alloys due to low costs, which can increase the TTs and possess high stability at higher temperature [3,4]. Likewise, the difficulties in fabrication of NiTi, NiTiHf alloys are also limited to simple geometries such as wires and rods, which can have limitations in many applications and is a major issue due to their lack of ability to fabricate complex parts. Additive manufacturing (AM) techniques have enhanced the fabrication of these complex geometries parts without any needs for tooling or machining via selective laser melting (SLM) technique [5-10].

Basically, during the additive manufacturing process, a laser melts down the powder and creates a melt pool of metal with the temperatures well above a couple of thousands of degrees in some cases. This leads to the formation of oxide on each layer and also the outer surfaces [11]. Oxidation affects the composition and also the thermomechanical behavior of the NiTi alloys, causing changes in the TTs or degradation of mechanical responses, for instance [12]. Therefore, it is required to analyze the high-temperature oxide formed on the NiTiHf alloys. To date, only Kim et al. [13] have evaluated the effect of the addition of Hf on the oxidation kinetic of a Ti-rich NiTi alloy during the heat treatment. It was found that the oxidation resistance of Ti-Ni-Hf was improved by adding 12 at. % Hf. The oxidation rate of Ti-Ni-Hf first followed by parabolic rate law and after some hours fitted with the linear rate law. The activation energy was calculated to be 324 KJ/mol in the parabolic stage, which was higher than that of those for NiTi alloys. The formation of a multi-layered oxide scale was observed, including an outer $TiO_2$ layer, a mixture of $TiO_2$ and $NiTiO_3$ layers, a layer of (Ti, Hf) oxides, and Hf rich layer, which had a significant effect on the enhancement of the oxidation resistance.

---

[1] Hdabbagh@rockets.utoledo.edu

Considering the development of AM techniques for NiTiHf alloys, no work has been carried on the evaluation of the oxidation resistance of AM NiTiHf in the literature compared to the conventionally made ones. The goal of this work is to analyze the oxidation behavior of AM NiTiHf alloys and compare the results with those of fabricated conventionally, at 500, 700, and 900°C.

## 2. MATERIALS AND METHODS

### 2.1 Material Preparation

As-cast $Ni_{50.4}Ti_{29.6}Hf_{20}$ (at. %) ingot was gas atomized by an electrode induction-melting gas atomization (EIGA) technique (TLS Technik GmbH (Bitterfield, Germany). Due to atomization with the inert gas, the EIGA technique produces a spherical powder with low impurity content in the range of 15-63 μm for NiTiHf parts to ensure followability and layer resolution. 4×4×10 mm coupons were fabricated by selective laser melting (SLM) method (Phenix Systems PXM) and then cut using wire electrical discharge machining (EDM). The SLM processing parameters for fabricating NiTiHf parts are shown in Table 1. Parameters are selected based on a previously published work in [14]. Conventionally made $Ni_{50.4}Ti_{29.6}Hf_{20}$ ingot used for comparison. Samples were cut from the same fabricated ingot that was atomized to make the powder for the SLM fabricated samples. More information on the fabrication and mechanical responses were presented elsewhere [14,15]. Hereafter, "AM" and "CON" were used for additively manufactured and conventionally made ingot, respectively.

**TABLE 1.** THE IMPLEMENTED PROCESSING PARAMETERS DURING SLM FABRICATION

| Type of Material | AM NiTiHf |
|---|---|
| Laser Power (P, W) | 150 |
| Scanning speed (v, mm/s) | 200 |
| Hatch Spacing (h, $\mu m$) | 80 |
| Layer Thickness ($\mu m$) | 30 |
| Energy Input (E, $\frac{J}{mm^3}$) | 313 |

Three small pieces of 40-200 mg AM and CON samples were cut and polished, then ultrasonically cleaned in acetone and dried to be used for thermogravimetric analysis (TGA) measurements. An SDT-Q600 instrument Combined with TGA was used for isothermal oxidation tests. The sample was maintained in an Alumina pan and heated up with at a rate of 10 °C/min in nitrogen gas and then maintained at 500, 700, and 900 °C with an airflow rate of 50 ml/min. After reaching the required temperature, the samples were oxidized in the air for 20 h, and the mass changes of the samples were measured. Keyence VHZ250R optical microscopy measured the surface area of the oxidized samples. The cross-section and the surface morphology of the oxidized samples were examined using a scanning electron microscope (SEM) equipped with an energy dispersive spectroscopy (EDS).

## 3. RESULTS AND DISCUSSION

The isothermal mass-gain versus time curves of the AM and CON NiTiHf alloys at a different temperature range of 500, 700, and 900 °C in the air up to 1200 min is depicted in Figure 1. As expected, increasing the temperature leads to an increase in the mass gain of all samples; however, the percentage of change is different. The overall increase in the mass of AM NiTiHf (solid lines) samples is lower than that of the CON ones (dashed lines), which shows better oxidation resistance of AM fabricated samples compared to those conventionally made NiTiHf. For AM alloys, the mass gain is very low at 500 and 700 °C, but then it significantly increases by raising the temperature to 900 °C. CON NiTiHf starts to oxidize very rapidly at 700 °C when compared to the other alloys at different temperatures.

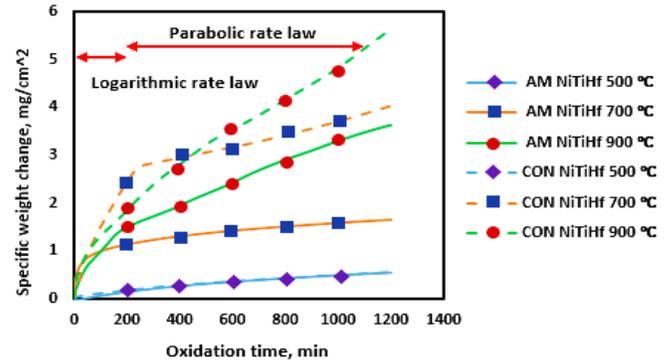

**FIGURE 1:** ISOTHERMAL OXIDATION CURVES OF AM AND CON NITIHF AT 500, 700, AND 900 °C

The oxidation kinetics of both AM and CON NiTiHf are not completely parabolic, as shown in Figure 1. In the initial stage of oxidation, all alloys follow logarithmic rate law based on equation 2, and then parabolic rate law (equation 1) which best fits with the isothermal oxidation curves [16]. The parabolic and logarithmic oxidation rate constants ($K_p$) and ($K_{log}$) are defined as follows:

$$(\frac{\Delta W}{A})^2 = K_p t \qquad (1)$$

$$\left(\frac{\Delta W}{A}\right) = K_{log} \log(t) \quad (2)$$

Where $\frac{\Delta W}{A}$ is the mass gain of the sample versus a unit of surface area, and t is the exposure time of oxidation. High correlation coefficient (R) values which are acquired from the linear regression for all samples. Parabolic rate constants of both alloys are shown in Table 2.

**TABLE 2.** CALCULATION OF PARABOLIC RATE CONSTANT FOR AM AND CON NITIHF AT 500, 700, AND 900 °C

| Alloys | Constants | Temperature (°C) | | |
|---|---|---|---|---|
| | | 500 | 700 | 900 |
| AM NiTiHf | $K_p\ (mg^2 cm^{-4} h^{-1})$ | 0.0208 | 0.0162 | 1.739 |
| | R | 0.99 | 0.99 | 0.99 |
| CON NiTiHf | $K_p\ (mg^2 cm^{-4} h^{-1})$ | 0.0210 | 2.702 | 3.07 |
| | R | 0.99 | 0.99 | 0.99 |

The Arrhenius plot of the parabolic rate constant is illustrated in Figure 2. Assuming that the activation energy is independent of the temperature, we can use the Arrhenius equation (shown in equation 3) to find the activation energies for the CON and AM NiTiHf.

$$k_r = A\,e^{-E_a/RT} \quad (3)$$

Where $k_r$ is the oxidation rate constant, A is a pre-exponential factor, $E_a$ is the activation energy, R is the gas constant, and T is the temperature [16].

According to the Table 2 and Figure 2, increasing the temperature leads to the increase in the oxidation constant rate and decrease the activation energy. The activation energy of AM and CON NiTiHf is found to be 60.634 KJ/mol and 91.454 KJ/mol, respectively. The activation energy for AM sample is shown to be lower than that of the CON one, which means that the reaction for AM NiTiHf changes more slowly than a reaction for CON NiTiHf [17].

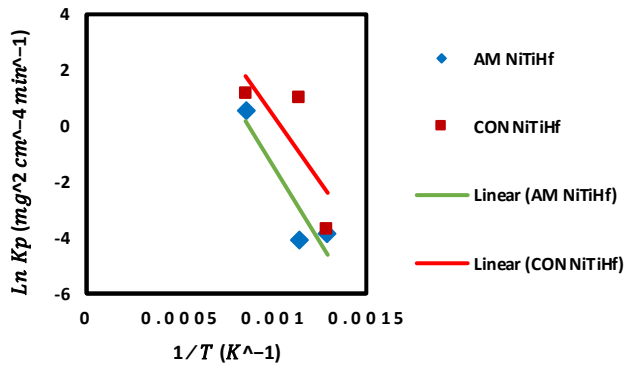

**FIGURE 2:** ARRHENUS PLOT OF PARABOLIC RATE CONSTANT FOR AM AND CON NITIHF

### 3.1 SEM results of the oxide layers

The oxide morphology of AM NiTiHf alloy exposed to isothermal flow at 900 °C for 75 h is shown in Figure 3. Based on the morphology of the oxide layers, at five different regions, the EDS analysis was done and reported in Table 3. The selected point of each analysis is shown in Figure 3.

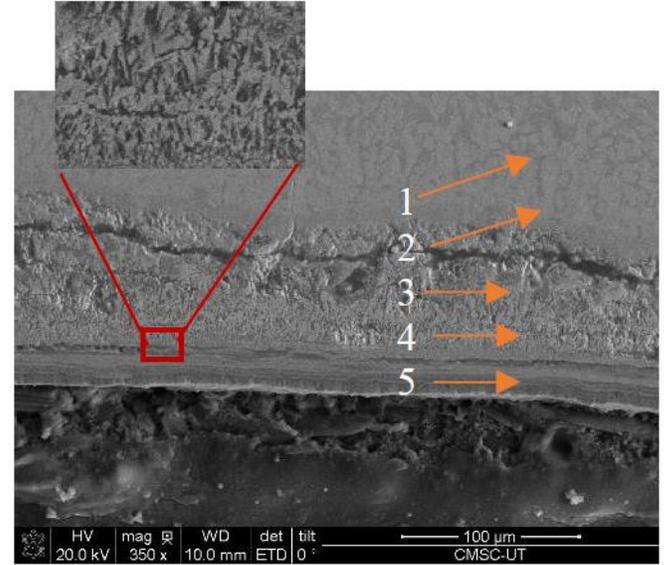

**FIGURE 3:** SEM MICROGRAPH OF CROSS-SECTION OF AM NITIHF AT 900 °C

Moreover, EDS elemental mapping of the same area is shown in Figure 4. X-ray mapping gives a better overview of the surface morphology with respect to the EDS analysis just performed on limited points. The mapping apparently confirms different layers. Also, the elemental rich or lean layers can be detected by this map, so that there is a thin layer of Ni-rich on the outer layer, while little or no Ni can be found in the intermediate layers. So, it confirms the NiTiO$_3$ formation on the outer layer.

**TABLE 3.** THE EDS ANALYSIS OF THE OXIDE LAYER OF AM NITIHF AT 900 °C

| Atomic% Points | Ti | Ni | Hf | O |
|---|---|---|---|---|
| 1 | 17.07 | 54.56 | 22.62 | 5.75 |
| 2 | 13.74 | 70.73 | 11.68 | 3.85 |
| 3 | 6.95 | 0 | 32.32 | 60.73 |
| 4 | 29.17 | 0.82 | 11.37 | 58.63 |
| 5 | 25 | 22.96 | 0 | 52.04 |

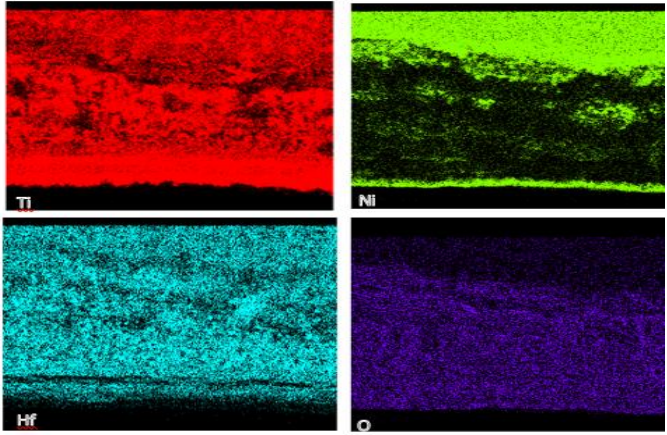

**FIGURE 4:** X-RAY MAPPING OF THE CROSS-SECTION OXIDE LAYER

Based on the EDS analysis, five distinct layers were detected. No Hf was detected in the dark outer layers, and it can be the oxide layers of $NiTiO_3$ + $TiO_2$, as Kim et al. reported in [13]. The light gray layer above the first layer is depleted from Ni (as it's shown in Figure 4), which can be resulted in a mixture layer of Hf, Ti oxide ($TiO_2$ + $HfO_2$). The cation size plays an important role in the diffusion rate. Hence, due to the smaller size of the Ti cation with respect to Hf cation, the Ti cation diffuses more upward to the outer oxide layers than Hf [18]. The needle-like structure is formed in the next layer (the inset of Figure 3). This Hf-rich layer approximately has allocated half of the whole oxide layers to itself, which is the thickest layer among all other oxide layers. Above the Hf-rich layer, a Ni-rich layer was formed, which has a significant Ni content up to 70 at. %. As the Ti and Hf diffused to the outer layer, the Ni-rich layer was formed above the oxide layers. The Ti depleted layer was formed as the last layer between the main NiTiHf matrix and the oxide layer.

Based on the oxide layer morphology, the needle-like Hf-rich layer is the thickest layer with approximately 50% of the oxide layer, while this number is about 25% for $TiO_2$ and $NiTiO_3$ oxide layers. So, the Hf-rich layer is a dominant factor in the kinetics of the oxidation. $TiO_2$ is the first oxide layer formed on the sample surface because of the lower Gibbs free energy with respect to $NiTiO_3$ and NiO [19,20]. After that, the Hf-rich layer started to form. At this stage, due to the difficulty of Ti diffusion through the $HfO_2$ [17,21], the Hf-rich layer blocks the upward diffusion of Ti to the oxide layers [13].

On the other hand, oxygen can easily diffuse into the $HfO_2$ layer and form Hf oxide. Therefore, the Hf-rich layer grew fast and became thick, while no significant change happened in $TiO_2$ and NiTiO layers thickness. This oxidation behavior can explain the two-stage oxidation kinetics; however, more investigation is needed to prove it. As it's shown in Figure 3, the crack was formed above the Hf-rich layer, which can be explained by the formation of a high brittle $HfO_2$ phase.

It should be noted that this analysis proves that the type of the oxide layers can be different from sample to sample based on the temperature of the oxidation and therefore, can be the source of variation in oxidation kinetic rates. Though further investigation is needed for the rest of the samples to assess the type of oxides layers and relate them to the temperature ranges. Also, as shown in Figure 1, rate of the oxidations is different for shorter duration (below 4 hours) for samples oxidized at 700 °C and 900 °C compared to the longer durations (20 hours). So, as the future study we also will consider the effect of time on the oxidation rate of NiTiHf to figure out the basis of differences in short duration oxidation at different temperatures.

## 4. CONCLUSION

The oxidation behavior of the AM NiTiHf parts fabricated by the SLM technique was studied and compared with the as-cast NiTiHf ingot. Both samples showed the two-stage oxidation behavior, the fast and short logarithmic behavior formed at the first stage, which was followed by the parabolic behavior for the rest. The oxidation rate of conventional samples is significantly more than the AM parts. Moreover, the morphology of the oxide layer for the AM part confirmed multi-layers oxidation, and it can explain the oxidation kinetics.